\documentclass[submission,copyright,creativecommons]{eptcs}
\providecommand{\event}{Eighth Workshop on Model-Based Testing 2013} % Name of the event you are submitting to
\usepackage{graphicx}
\usepackage{breakurl}
\usepackage[pdftex]{color}

\usepackage{amssymb,amsmath} % amsmath nÃtig für qed und wegen misplaces alignment tab

\title{Towards the Usage of MBT at ETSI}
\author{Jens Grabowski
\institute{University of G\"ottingen, Germany}
\email{grabowski@informatik.uni-goettingen.de}
\and
Victor Kuliamin
\institute{ISP RAS, Russia}
\email{kuliamin@ispras.ru}
\and
Alain-Georges Vouffo Feudjio
\institute{Thales, Germany}
\email{alain-georges.vouffofeudjio@thalesgroup.com}
\and
Antal Wu-Hen-Chang
\institute{Ericsson, Hungary}
\email{antal.wu-hen-chang@ericsson.com}
\and
Milan Zoric
\institute{ETSI, France}
\email{milan.zoric@etsi.org}
}
\def\titlerunning{Towards the Usage of MBT at ETSI}
\def\authorrunning{Grabowski et al.}
\begin{document}
\maketitle

\begin{abstract}
In 2012 the Specialists Task Force (STF) 442 appointed by the European
Telcommunication Standards Institute (ETSI) explored the possibilities
of using Model Based Testing (MBT) for test development in
standardization. STF~442 performed two case studies and developed an
MBT-methodology for ETSI. The case studies were based on the
ETSI-standards GeoNetworking protocol (ETSI TS~102 636) and the
Diameter-based Rx protocol (ETSI TS~129~214). Models have been
developed for parts of both standards and four different MBT-tools
have been employed for generating test cases from the models. The case
studies were successful in the sense that all the tools were able to
produce the test suites having the same test adequacy as the
corresponding manually developed conformance test suites. The
MBT-methodology developed by STF~442 is based on the experiences with
the case studies. It focusses on integrating MBT into the
sophisticated standardization process at ETSI. This paper summarizes
the results of the STF~442 work.
\end{abstract}

\vspace{-0.3cm}
\section{Introduction}
\label{Introduction}

Driven by technological advances and an ever-growing need for software 
and systems quality improvements, MBT has matured in the last decade from 
a topic of research into an industrial technology. MBT has been successfully 
used for the automatic generation of test documentation and test scripts in 
a wide range of application areas including information and communication 
technology, embedded systems and medical software. This trend is reflected 
by the availability of various commercial tools and increasing efforts in 
MBT-related standardization. The utilization of MBT in industry show 
significant gains in productivity, in particular due to savings in 
the test maintenance phase.

 In 2010, the ETSI Technical Committee (TC) on Methods for Testing and 
Specification (MTS) published a first ETSI standard on MBT 
(ES~202~951)~\cite{rqmn} as the result of a joint effort of different 
stakeholders at ETSI including MBT tool vendors, major users, service 
providers, and research institutes. In order to enable the use of this 
technology at ETSI, the applicability of MBT in ETSI processes has to be 
shown and methodology guidelines for applying MBT in the context of 
standardized test development are needed. For this purpose ETSI TC MTS 
started in~2012 STF~442. STF~442 consists of five experts from industry 
and academia with~30 working days each. The work was conducted from 
February 2012 to December 2012. STF~442 performed two case studies from 
the ETSI domains Intelligent Transportation Systems (ITS) and Universal 
Mobile Telecommunications System (UMTS) and used the gained experience 
for developing ETSI MBT methodology guidelines.

 In the following, we present the case studies, describe the methodology 
and discuss problems encountered when applying MBT in the case studies.

\vspace{-0.3cm}
\section{Case Studies}
\label{Case Studies}

The following four MBT tools have been used for the case studies:

\begin{itemize}

\item \textbf{Conformiq Designer} is the MBT tool of Conformiq
Inc. \cite{conformiq}. Conformiq models are written in a combination
of Java code and UML statecharts, i.e., in the Conformiq Modeling
Language (QML). The models describe the expected external behavior of
the System Under Test (SUT). Java code is used to describe the data
processing of the SUT, to declare data types and classes, to express
arithmetics and conditional rules as well as others. UML statecharts
are used to capture high-level control flow and life cycle of
objects. The core of Conformiq Designer is its semantics driven,
symbolic execution based test generation algorithm. The algorithm
traverses a part of the (usually infinite) state space of the system
model. The test generation heuristics that Conformiq Designer uses
realize various well-known test generation strategies, e.g.,
requirements coverage, transition coverage, branch coverage, atomic
condition coverage, and boundary value analysis.

\item \textbf{Microsoft Spec Explorer} for VisualStudio 2010 is a
Microsoft MBT tool \cite{microsoft}. Spec Explorer uses state-oriented
model programs that are coded in C\#. Test generation is performed by
exploring the state space of the system model and recording the
traces. These traces are transformed into test cases. The main
technique for dealing with state space explosion provided by Spec
Explorer is scenario-based slicing. A scenario limits the potential
executions of the model state graph, while preserving the test
oracle and other semantic constraints from the system model. Slicing
scenarios along with test data used as input for model operations are
defined in the scripting language Cord.

\item \textbf{Sepp.med MBTsuite} is the MBT framework from the
sepp.med GmbH \cite{seppmed}. For applying MBTsuite, a graphical model
of the SUT has to be provided. In our case studies, UML state and
activity diagrams have been used.  MBTsuite excutes models and
transforms the execution traces into test cases.  Apart from full path
coverage, other generation strategies are available (e.g. guided
generation, random generation). If defined in the model, guard
conditions and priorities are taken into account at execution
time. Thus, only logically consistent execution traces are obtained
and processed into test cases. It is possible to filter the execution
traces prior to test case generation using several built-in heuristics
like, e.g., node coverage, edge coverage, requirement coverage, but
also heuristics based on test management information (costs,
duration).

\item \textbf{Fraunhofer MDTester} is an academic MBT tool developed
by the Fraunhofer FOKUS competence center MOTION \cite{fokus}.
MDTester is part of Fokus!MBT, a flexible and extensible test modeling
environment based on the UML Testing Profile (UTP), which
facilitates the development of model-based testing scenarios for
heterogeneous application domains. MDTester is a modeling tool that
guides the development of UTP models. UTP models are test models and
not system models, i.e., they include tester knowledge like, e.g.,
setting of test verdicts, knowledge about test components, or default
behavior.  For modeling, MDTester provides the following diagrams
types: test requirements diagram (based on class diagram), test
architecture diagram (based on class diagram), test data diagram
(based on class diagram), test architecture diagram, test behavior
diagram (based on sequence and activity diagrams).
\end{itemize}

The case studies were based on ITS and UMTS protocols standardized by
ETSI. In addition, STF~442 conducted the academic example of a simple
automated teller machine to gain experience with the tools.

For the ITS-based case study, conformance tests for the location service 
functionality of the GeoNetworking protocol (ETSI TS~102~636)~\cite{geonet} 
have been generated from previously developed models. The GeoNetworking 
protocol belongs to the ITS network layer. The location service 
functionality is used to discover units with certain addresses and 
to maintain data on their geographical location.

The Rx interface (ETSI TS~129~214)~\cite{rx} of UMTS provides the base 
for the second case study. The Rx interface supports the transfer of 
session information and policy/charging data between Application Function 
and Policy/Charging Rules Function on top of the Diameter protocol.

In both case studies, the modeled behavior of the System Under Test (SUT) 
can be described with approximately 12 control states and a slightly 
higher number of transitions between them. However, the main complexity 
of the SUT-model behavior is related to data stored and used during 
operation. For the GeoNetworking case study, this data refers to 
addresses and geographical locations; whereas session settings and policy 
rules are most important the behavior of the Rx interface case study.

%%% \textbf{An example of (if we have space )}

Two different approaches have been used for modeling. The first approach 
started from the manually developed test purposes~\cite{geonettp,rxtp} 
and resulted in SUT-models sufficient to cover all the test purposes, 
meanwhile adding some more details from standard requirements. The second 
approach was based on the requirements in the base standard. 
The constructed SUT model tried to reflect all of them in their behavior. 
Both approaches were successful in a sense that the models were suitable 
for test generation. 

In spite of the fact the different tools use different formalisms as input 
for SUT models and provide different means to control test generation, all 
tools managed to generate test suites that cover almost the manually 
developed test purposes. Thus, from a technical point of view, modern 
MBT tools are able to support test development in standardization.

The case studies are documented in~\cite{mbtcs}. The report includes 
detailed descriptions of the SUT behavior and the models, a discussion 
of modeling approaches, the generated tests, and overall evaluation. % of all case studies.

\vspace{-0.3cm}
\section{Methodology Guidelines}
\vspace{-0.2cm}

\begin{figure}[t]
\centering
\vspace{-0.5cm}
\includegraphics[width=0.95\textwidth]{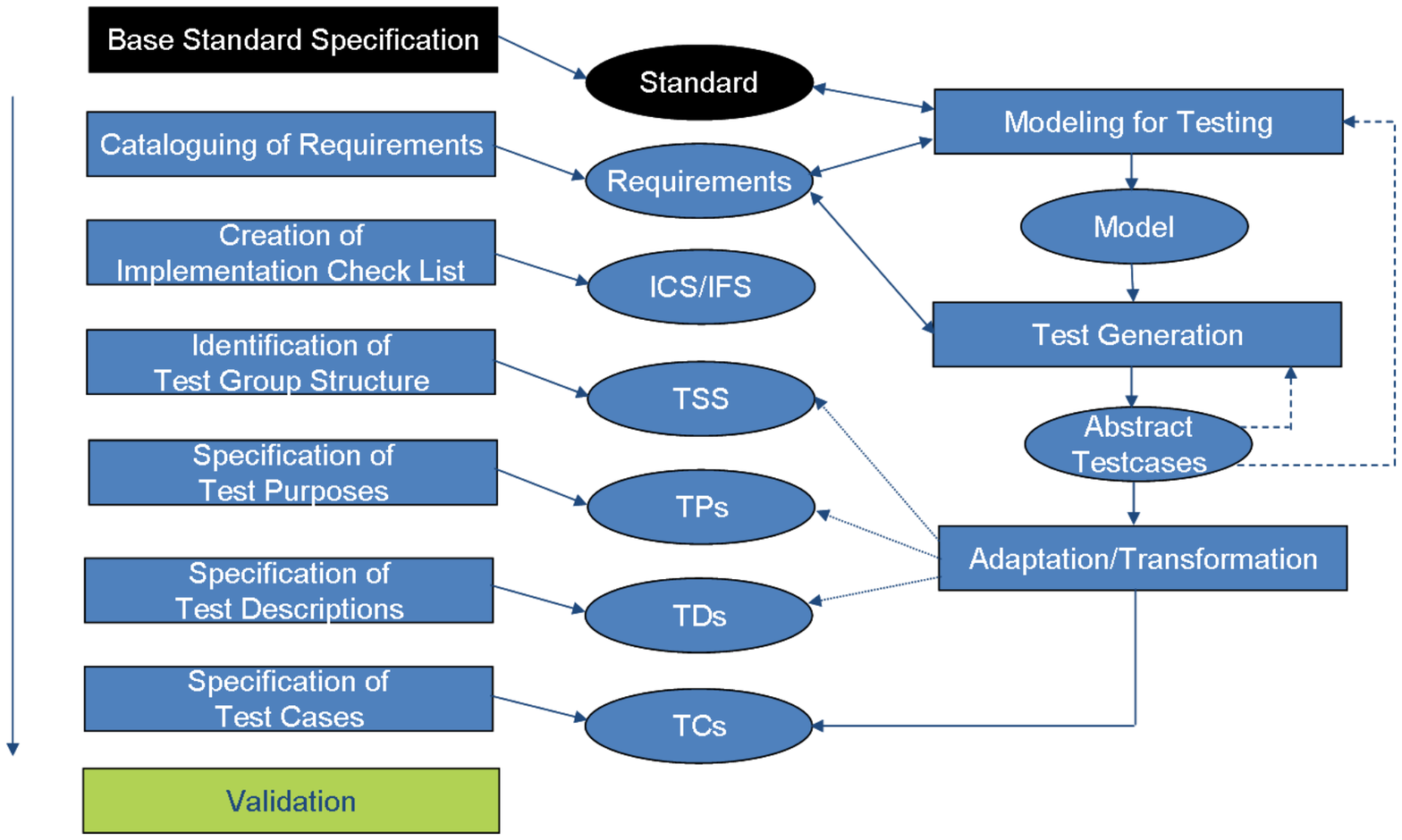}
\vspace{-12.6cm}
\caption{Using MBT within the ETSI test development process}
\vspace{-0.3cm}
\label{process}
\end{figure}

The second goal of the STF work was the development of methodology 
guidelines for an MBT-based development of conformance tests at 
ETSI~\cite{mbtmthg}. ETSI has a very sophisticated test development 
procedure shown on the left side of Figure~\ref{process}. Test 
development starts with the identification of requirements followed 
by the creation of Implementation Conformance Statement (ICS) and 
Interoperable Function Statement (IFS). ICS/IFS define implementation 
options for a standard. In the testing process, they are used for test 
case selection. The ICS/IFS creation is followed by the specification 
of the test suite structure, which in most cases arises from 
the functionality to be tested. Afterwards, high-level test 
descriptions, i.e., test purposes, are stepwise refined leading to 
the test cases, which are finally validated. The test development steps 
lead to documents represented by the ellipses in the middle of 
Figure~\ref{process}.

The integration of MBT in the ETSI process is shown on the right side 
of Figure~\ref{process}. The modeling for testing is based on standard 
and requirements. If possible, implementation options (i.e., ICS/IFS) 
are considered in modeling. The modeling process can be seen as 
an additional validation step for the standard, the requirements and 
the implementation options. Problems in modeling may identify ambiguities 
in the standard or untestable requirements. The model serves as input 
for the test generation. Problems identified during test generation 
or in the generated tests may identify problematic requirements 
or require adaptations in the SUT model. For integrating MBT into 
the ETSI test development process, documents describing test suite 
structure, test purposes, test descriptions and test cases have to 
be generated.

Even though this embedding of MBT into the ETSI test development 
process looks straightforward, several issues need to be solved 
before MBT can improve the existing process. A main problem is 
the maintenance and consistency of model and test documents. 
On the one hand, MBT only requires maintenance and further development 
of models while test cases are generated and not manually developed. 
On the other hand, each test case is an asset and its implementation 
can be very costly. Reviews and discussions are therefore mainly 
based on individual test descriptions and not on models. Another 
issue is the selection of a modeling language. Even though all MBT 
tools used for the case studies allow state-oriented modeling, 
the input languages differ considerably. A pragmatic solution to 
this problem may include the standardization of an ETSI modeling 
language.

In addition to issues regarding the test development process, 
the ETSI MBT methodology guidelines also offer guidance for 
identification and modeling of requirements, establishing 
traceability from models to standard requirements, choosing model 
scope and abstraction level, selecting test coverage criteria, 
improving maintainability and parameterization of generated tests, 
as well as assessing the quality of models and tests.

\vspace{-0.2cm}
\section{Summary and Conclusions}

STF~442 has successfully applied MBT to generate conformance tests 
for two ETSI protocols. Both case studies have been performed with 
all tools. All tools were able to generate to test suites having 
an adequacy level comparable with manually designed tests.  Based on 
the case studies, ETSI MBT methodology guidelines have been developed. 
The methodology guidelines focus on integrating MBT into 
the standardization process at ETSI. Some challenges have been 
identified during the STF work:

\begin{itemize}
\item An efficient usage of MBT in standardization requires 
  significant expertise in several areas, like e.g., the domain 
  of the SUT, modeling, MBT tool application, and test development. 
  Experts experienced in all areas are difficult to find.

\item There exists an abstraction gap between automatically generated 
  and manually specified test cases. Manually test cases are usually 
  more maintainable and can be subject of a review. By considering 
  parameterization, manually developed test cases allow an easy 
  adaptation to different implementations of a standard. Solving this 
  issue can be seen as a requirement for future MBT tools.

\item The conformance test development process at ETSI is tightly 
  intertwined with test suite maintenance issues and with handling 
  each test case as a separate artifact. Test cases are designed 
  individually and are subject of discussions and reviews. In contrast 
  to the ETSI process, one of the main MBT advantages is the transfer 
  of all maintenance work to the modeling, while tests are considered 
  to be generated automatically as often as needed, i.e., maintenance 
  of automatically generated tests is not necessary. For ETSI, taking 
  full advantage of MBT may require new processes changing from 
  the test case centric development to model standardization and maintenance.

\end{itemize}

\subsection*{Acknowledgements}
The authors thank ETSI TC MTS for supporting the work presented in this paper.

\vspace{-0.3cm}
\bibliographystyle{eptcs}
\bibliography{mbt2013}
\end{document}